\baselineskip = 18pt
\nopagenumbers
\font\large = cmr10 scaled 1200
\hsize = 6.5 true in
\vsize = 8.5 true in
\parskip = -1pt 
\def\reallynarrow{\advance\leftskip by 3\parindent \advance \rightskip by 3\parindent}
\def\sc{\scriptscriptstyle}
\def\HI{\tau_{\sc\rm GP}^{\sc\rm H \,I}}
\def\HeI{\tau_{\sc\rm GP}^{\sc\rm He \,I}}
\def\HeII{\tau_{\sc\rm GP}^{\sc\rm He \,II}}
\def\apj#1  #2 #3 {Astrophys. J. #1,  #2 (#3)}
\def\mnras #1 #2 #3 {MNRAS #1, #2 (#3)}
\def\etal{\it et al.\/}
\def\la{\mathrel{\mathchoice {\vcenter{\offinterlineskip\halign{\hfil
$\displaystyle##$\hfil\cr<\cr\sim\cr}}}
{\vcenter{\offinterlineskip\halign{\hfil$\textstyle##$\hfil\cr<\cr\sim\cr}}}
{\vcenter{\offinterlineskip\halign{\hfil$\scriptstyle##$\hfil\cr<\cr\sim\cr}}}
{\vcenter{\offinterlineskip\halign{\hfil$\scriptscriptstyle##$\hfil\cr<\cr\sim\cr}}}}}
\def\ga{\mathrel{\mathchoice {\vcenter{\offinterlineskip\halign{\hfil
$\displaystyle##$\hfil\cr>\cr\sim\cr}}}
{\vcenter{\offinterlineskip\halign{\hfil$\textstyle##$\hfil\cr>\cr\sim\cr}}}
{\vcenter{\offinterlineskip\halign{\hfil$\scriptstyle##$\hfil\cr>\cr\sim\cr}}}
{\vcenter{\offinterlineskip\halign{\hfil$\scriptscriptstyle##$\hfil\cr>\cr\sim\cr}}}}}


\vfill

\centerline {\large A New Astrophysical Constraint on Radiatively 
 Decaying Neutrinos } 

\bigskip

\centerline {\bf Shiv  K.  Sethi} 


\centerline{ Inter University Centre for Astronomy and Astrophysics }

\centerline{ Post Bag 4, Ganeshkhind, }

\centerline{ Pune 411007 }




\vfill

\beginsection \centerline {Abstract}

 We calculate constraints on radiatively decaying neutrinos from the recent detection of  singly ionized helium in the diffuse intergalactic medium (IGM) at $z \simeq 3.3$. We consider a model in which neutrinos predominantly decay into invisible relativistic particles with a rate $\tau^{-1}$, and with a small branching ratio into the radiative mode.  To satisfy the observation of singly ionized helium, which puts a lower bound on the number density of singly ionized helium in the IGM, we show that: for $\tau \ga 10^{18} \,{\rm sec}(1 \,{\rm eV}/m_\nu^2)$, transition moment of neutrinos $\mu_{12}$  is constrained to be   $\le 4 \hbox{--}8 \times 10^{-17} \,\mu_{\sc B}$ for $ 110 \,{\rm eV} \la m_\nu \la 10 \,{\rm keV}$.
We compare this bound with other astrophysical and cosmological bounds on radiatively decaying neutrinos.


\bigskip

\noindent PACS numbers: 98.60.Hj, 98.80.Cq, 98.70.Vc, 98.80.bq 
 
\vfill\eject 

\pageno =1
\footline={\hss\tenrm\folio\hss}


Radiatively decaying neutrinos have found widespread application in
cosmology and astrophysics [1-3]. On the other hand, a formidable array of
constraints exist on radiative lifetime of radiatively
decaying neutrinos, from studies of  various disparate phenomena in
cosmology and astrophysics [3-8]. In this paper, we point out another
constraint which comes from the existence of  primordial
elements (hydrogen and helium) in the diffuse IGM
 at high redshifts. The search of primordial elements in the diffuse IGM   is crucial to understanding the universe at early epochs (Gunn-Peterson (GP) test [9]). So far there is no definite evidence of the presence of either neutral hydrogen or neutral helium in  diffuse IGM [10,11].   Last year, first detection of  singly ionized helium in diffuse IGM  was reported by Jakobsen {\etal.} [12]. 
Jakobsen {\etal}\ detected  strong absorption from singly ionized helium along the line of sight of  quasar Q0302-003 at $z \simeq 3.3$. Their 90\% upper bound on the continuum optical depth from singly ionized helium was $\HeII \ge 1.7$. As there are only upper bounds on $\HI$ and $\HeI$, the presence of singly ionized helium in the IGM puts stringent constraints on the magnitude  and spectrum of  homogeneous sources of photoionization [13]. Radiatively decaying neutrinos is one such homogenous source of photoionization. In this scenario, a massive neutrino ($\nu_h$) decays into a photon ($\gamma$) and a light neutrino ($\nu_\ell$) [14,15]. If $m_{\nu_\ell} \ll m_{\nu_h}$---the case we consider---the photon and the light neutrino equally share the rest energy of $\nu_h$. We assume the heavy neutrino to be tau neutrino ($\nu_\tau$), while the light neutrino can be either $\nu_\mu$ or $\nu_e$. To ionize singly ionized helium the mass of tau neutrino must be greater than~108.8~eV (the ionization potential of singly ionized helium is~54.4~eV). The closure density constraint restricts the mass of a stable neutrino (or a neutrino which decays with a rate less than $t_0^{-1}$, $t_0$ being the present age of the universe) to be less than~$91\rm h^2\,eV$. This constraint forces the rate of neutrino decay to be greater than~$t_0^{-1}$. As  a radiative decay lifetime of less than~$t_0$ is forbidden by several constraints, the neutrinos must predominantly decay into particles which are invisible (majoron, neutrinos etc.) with decay rate $\tau^{-1} \gg t_0^{-1}$, with a small branching ratio $B$ for the radiative mode (radiative lifetime $\tau_\gamma = \tau /B$) .  

In the presence of decaying neutrino, the thermal history of the
universe can change considerably. First, as soon as the neutrinos
become nonrelativistic the universe becomes matter dominated ($z =
z_1$). The universe remain matter dominated, with massive neutrinos
dominating the energy density, until the time the neutrinos decay ($z
= z_d$) into relativistic decay products. (For simplicity we assume
that all relativistic products of decaying neutrinos remain
relativistic upto the present epoch.) Thus the epoch of neutrino decay is followed by a period of radiation domination in the universe. Depending on the mass and  lifetime of decaying neutrinos and the energy density of other nonrelativistic matter ($\Omega_{NR}$) in the universe, another period of matter domination could follow ($z = z_2$). In the foregoing we have assumed the neutrinos to decay instantanously, which would result in a sudden change in matter and radiation domination. It is a good assumption for calculating the time-redshift relationship and for computing the age of the universe in the presence of decaying neutrinos [16].
 In this paper we assume the universe to be spatially flat i.e., $
\Omega_0 \simeq \Omega_\nu + \Omega_{NR} = 1$, where $\Omega_{NR}$ is
the energy density due to all other non-relativistic matter. 
 
Helium is the second most dominant primordial element ( 8 \% by number). In a uniform IGM, the proper density of helium is 
$$ n_{\rm \sc He} = 6.8 \times 10^{-7} \Omega_{IGM} h^2 \,\, \rm cm^{-3}.\eqno(1) $$
Throughout this paper we take $\Omega_{IGM} = 0.05$ and $h = 1/2$.
For a  uniform IGM, any resonance line at wavelength $\lambda$ and oscillator strength f, of singly ionized helium  with a  proper number density $n_{\rm \sc He \, II}(z)$, will produce an optical depth [17,18] 
$$ \HeII = {\pi^2 \over m_e c} f H^{-1}(z) \lambda n_{\rm \sc He \, II}(z). \eqno (2)$$
 The dominant scattering is at wavelength   304 $\rm{\AA}$. Using this, $\HeII >1.7$ at $z = 3.3$ translates into a lower bound on the fraction of  singly ionized helium, $ y_2 > 1.2 \times 10^{-3}$, where $ y_2 = n_{\rm \sc He \, II}/n_{\rm \sc He}$. Using that lower limit on $y_2$ is much less than unity,  equilibrium  between recombination and ionization processes can be assumed, which allows one to express $y_2$ as
$$ y_2 \simeq  0.026\times 10^{-10}{(1+z)^3 \over n_\gamma(z)}. \eqno(3)$$ 
Here 
$$ n_\gamma(z) = {\epsilon_0^3 \over \pi^2} \int\limits_{\epsilon_0}^{m_\nu/2}\!\!{dk \over k} f_d(k,t) \exp(-\tau(z,z_e,k) \quad \rm cm^{-3}; \eqno(4)$$ 
$\epsilon_0 = 54.4 \, \rm eV $ is the ionization potential of singly ionized helium;  we have assumed the temperature of IGM $T = 1.5 \times 10^4  \, \rm K $; $f_d(k,t)$ is the distribution function of decay photons; $\tau(z,z_e,k)$ is the optical depth (to be described below) suffered by a photon which is emitted at $z_e$ and   observed at  $z$ with a wave number $k$. In the limit of small energy transfer between electrons and decay photons (a valid assumption for $z_d \la 10^4)$, the distribution function of photons can be written as [1,8]
$$ f_d(k,t) = { B \pi^2 n_\nu(\tau) \over n (m_\nu/2)^{1/n}k^{3-1/n}} \left \lbrack {\tau \over t} \right \rbrack ^{3n-1} \exp \left \lbrack - \left(t/\tau \right) \left (2k/m_\nu \right) ^{1/n} \right \rbrack \Theta(m_\nu/2 - k). \eqno(5)$$
Here n = 1/2, 2/3 for radiation-dominated and matter-dominated epochs respectively.  

Before discussing the results of our analysis, we briefly review some of the other bounds on radiatively decaying neutrinos. If  neutrinos have a  coupling to photons then, during a supernovae explosion, a photon flux coincidental with the neutrino flux  from the supernovae should be observed. No such flux was observed during supernovae SN1987A. This has been used to constrain the parameter space of radiatively decaying neutrino [6]:
$$  \tau \,m_\nu \geq  2\times 10^{19} B \,\,{\rm eV \,sec}, \eqno(6)$$ 

for neutrino masses between 100 eV and a few MeV. Other major astrophysical constraint comes from the study of red giant stars. The dominant energy loss mechanism in red giants is the decay of plasmons, formed in the core of the star.  The rate of plasmon decay is enhanced if neutrinos have a magnetic moment. This can lead to  precipitous cooling of the star, which is in conflict with  known stellar evolution time scales. A study of such a process constrains the neutrino transition moment $\mu_{12} \la 10^{-12} \mu_{\sc B}$, which translates to [7]
$$ \tau \ga 2.1 \times 10^{22} B \, {\rm sec} \,\left({m_\nu \over 1\, {\rm eV}} \right)^{-3}, \qquad\quad\hbox{for $m_\nu \le 10 \,\rm keV$.} \eqno(6)$$

An interesting bound on trasition moment of radiatively decaying  Dirac neutrinos can be obtained if the universe had a primordial magnetic field. The presence of magnetic field can cause the right-handed neutrino to be in theraml equilibrium with other relativistic particles in the universe at the time of nucleosynthesis, thereby adding an extra   relativistic species, which  is unaccetable. From such an analysis, a lower bound on trasition moment $ \mu_{12} \le 5 \times 10^{-14} (30 \, {\rm eV} /m_\nu)^{4/3} \, \mu_B$ can be obtained [19]; or in terms of neutrino lifetime
$$ \tau \ge 2.5 \times 10^{22} \left ({30 \, {\rm eV} \over m_\nu } \right ) ^{4/3} \,\, \rm sec.$$
However, as there is no evidence of the  existence of primordial magnetic field, this constraint may not apply.

Another stringent constraint on radiatively decaying neutrinos comes from the planckian nature of CBR. Photons emitted by decaying neutrinos can heat the electrons in the IGM;  hot electrons can transfer this energy to CBR photons. If this process takes place at redshifts $ \la 10^5$, the spectrum of CBR is altered [20]. As CBR is known to be a planckian to a very high accuracy [21], the decay of neutrinos into photon mode is severly restricted~[8] 
$$ B \,\left({m_\nu \over 1\, {\rm eV}} \right)^2 \la 4.1 \times 10^2. \eqno(7)$$ 

Finally, the decay  photons constitute a diffuse background at wavelengths $ \la m_\nu/2$. In recent years,  rocket-borne experiments have placed upper bounds on diffuse, background photon flux  in  ultra-violet (UV), a wavelength range in which radiatively decaying neutrinos contribute most significantly [22]; in addition there exist  several  bounds  on the extragalctic hydrogen-ionizing flux  from astrophysical observations [23]. 

In this paper we point out that radiatively decaying neutrinos with masses above 108.8~eV are subject to an additional constraint: the existence of singly ionized helium in the IGM. From eq.(3), it  is clear that if $  n_\gamma(3.3) \ga 10^{-7}$, observed  bound on $\HeII$ would be violated. To calculate $  n_\gamma(z)$, sources of absorption   must be identified. We consider two sources of absorption: diffuse IGM and lyman-$\alpha$ systems. Optical depth due to diffuse IGM is
$$  \tau_{\sc \rm diffuse}(z,z_e,k) \simeq \int_{z_e}^z dz \left \vert{dt \over dz} \right\vert \,n_{\sc\rm He}\sigma_{\sc\rm HeII}(k)y_2, \eqno(8)$$
where  $\sigma_{\sc\rm HeII}(k)$ is  the photoionization cross section from the ground state [24]. $\tau_{\rm cloud}$, the avarage attenuation of photon flux due to poisson-distributed clouds, is given by
$$\tau_{\rm cloud}(z,z_e,k) \simeq  \int_{z_e}^z \int_0^\infty dz \, dN_{\sc\rm HI}\, {\cal P}(N_{\sc \rm HI},z)\{1-\exp \lbrack  - N_{\sc\rm HeII}\sigma_{\sc\rm HeII}(k) \rbrack\}. \eqno(9)$$
Here $N_{\sc\rm HI}$ and $N_{\sc\rm HeII}$ correspond to column densities of neutral hydrogen and singly ionized helium in the clouds and  ${\cal P}(N_{\sc\rm HI},z)$ is the number density of clouds in a given column density and redshift interval. In ionization equilibrium, $N_{\sc\rm HeII}$ can be inferred from known column densities of neutral hydrogen $N_{\sc\rm HI}$
$$ N_{\sc\rm HeII} = N_{\sc\rm HI} \times { n_\gamma^{\sc\rm HI} \over n_\gamma^{\sc\rm HeII}} \quad \rm cm^{-2}; \eqno(10) $$
here $ n_\gamma^{\sc\rm HI} $ is the effective number of photons which ionize neutral hydrogen. From 'proximity effect' [25], $ n_\gamma^{\sc\rm HI} $ is known to approximately constant for $2 \la z \la 4$ with a lower limit  $ n_\gamma^{\sc\rm HI}  \ga 6.2 \times 10^{-5} \rm cm^{-3}$. For some of the paramter space for which $\HeII \ga 1$, the value of $ n_\gamma^{\sc\rm HI} $ from decaying neutrinos is less than the lower limit from 'proximity effect'. Hence we include the contribution from 'proximity effect' in  $ n_\gamma^{\sc\rm HI} $ . It is necessary to include this additional source of photons to correctly determine the optical depth (eq.(10)).    
For ${\cal P}(N,z)$, we take models A1 and A2 of Miralda-Escud\'e and Ostriker [18], in the redshift range $ 1 \le z \le 5$.  

Our results are shown in figs. 1--4, along with other constraints.
To gauge the effect of absorption, results in case of no absorption
are also shown. In some range of masses, constraints on $B$ from the
detection of singly ionized helium can be a few { \it orders \/} more
stringent than any other constraint. Our results can be qualitatively
understood. The photon flux at any energy E is determined by the the
shape of decay photon spectrum (eq(5)). At any given redshift, if the
exponent in photon spectrum can be neglected for  photon energies
$\ga  54.4 \,\rm eV$
then photon flux  at these energies
$${\cal F}(E) \propto E^3 \times f_d(E) \simeq E^{3-1/n}. $$
Thus photon spectrum is extremely hard (for comparsion, ${\cal F}
\propto E^{-\alpha}$, $0.7 \le \alpha \le 1.5$, for a photon background
dominated by quasars). The bounds on $B$ are most stringent in this
region. As $m_\nu$ is decreased  for a given $\tau$, the exponent in
decay photon spectrum begins to be important. This leads to a decrease
in the number of  photons above energy  54.4 eV as well as to an increase in absorption (eq(10)), 
because the photon spectrum becomes soft for energies $\simeq 54.4 \,
\rm eV$. And so larger values of $B$ are required to ionize the singly
ionized helium. Our results are valid for $10^{12}\, {\rm sec} \le
\tau \le 10^{16}\, \rm sec$. It is because for $\tau$ much less than
$10^{12}$, the assumption that interaction between photons and
electrons can be neglected in determining the spectrum of the decay
photon breaks down. For $\tau \ge 10^{16} \, \rm sec$, neutrino masses
above 108.8 eV violate the closure density constraints. (All the region shown in figs (1)--(4) is compatible with closure density bounds.)
 The bounds on radiative lifetime can be translated into bounds on magnetic moment $\mu_\nu$ of radiatively decaying neutrinos. For Dirac neutrinos, radiative lifetime $\tau_\gamma$ (= $\tau/B$) is related to neutrino  transition  moment~$\mu_{12}$ as [26] 
$$ \tau_\gamma = 10^{23}\left({30 \, {\rm eV} \over m_\nu}\right)^3 \left ( {10^{-14} \mu_{\sc B} \over \mu_{12}} \right)^2 \,\,\rm sec.$$

The bounds on $\mu_{12}$ from the existence of singly ionized helium in the IGM can be summarized, for model A2 for absorption from lyman-$\alpha$ systems, as:
 {\item{(1)} For $\tau > 1.6 \times 10^{18} \, {\rm sec} (1 \,{\rm eV}/m_\nu^2)$,
$$\mu_{12}\la 4\hbox{--}8 \times 10^{-17} \, \mu_{\sc B},$$
in the mass range $ 110 \,{\rm eV} \la m_\nu \la 10 \,{\rm keV}$.
\item{(2)} For $ 1.6 \times 10^{18} \,{\rm sec}(1 \, {\rm eV}/m_\nu^2) \ga \tau \la 5\times 10^{17} \,{\rm sec}(1 \, {\rm eV}/m_\nu^2)$,
 
$$\mu_{12} \la 8\hbox{--}100\times 10^{-17} \, \mu_{\sc B},$$

in the mass range $ 110 \,{\rm eV} \la m_\nu \la 10 \,{\rm keV}$. }

\noindent  For $\tau \la 5\times 10^{17} \,{\rm sec}(1 \,{\rm eV}/m_\nu^2)$, other bounds become more significant (figs. 1--4).

Our bounds on $\mu_{12}$ are not restrictive enough to constrain the standard Wienberg-Salam model. However, several extensions of the standard model like the left-right symmetric model, charged-Higgs model, and broken R-parity supersymmetric model can give large transition moments (for a recent review see [27], [28]), and therefore would be constrained by the bounds we derive. As the most stringent constraints are obtained  in  the range  $ \tau \ga  1.6 \times 10^{18} \, {\rm sec} (1 \, {\rm eV}/m_\nu^2)$, our results can be used to put an upper bound on $\tau$. It should be pointed out that the though these values of $\tau$ are not constrained by the closure density constraints, age and structure formation constraints may rule out much of the range of $\tau$ in which most stringent constraints on $\mu_{12}$ are obtained ([29], [30]).

We now discuss various sources of errors in our estimates from uncertainties in input parameters. 
(As GP tests are quite sensitive and model independent, our analysis is  free from  any  errors of modelling.) First, it has been pointed out that a large contribution to Jakobsen et al.'s bound of $\HeII \ge 1.7$ might come from line blankating [13]. But the presence of a foreground quasar along the line of sight to Q302-003 might reduce the optical depth due to line blankating quite significantly [31]. In any case, these uncertainties should not change  bounds on $B$  by more than an order. (It should however be pointed out that even in case most of the contribution to optical depth observed by Jakobsen {\etal} comes from line blankating, equally stringent bounds on radiatively decaying neutrinos can be obtained. If line blankating causes most of the optical depth, one requires $S_L \equiv n_\gamma^{\sc\rm HI}/ n_\gamma^{\sc\rm HeII} \ga 65 $. This puts strong constraints on the spectrum of radiatively decaying neutrinos.) Other input parameters like $\Omega_{IGM}$, $T$, and $h$ can change our estimate by a factor of few.  Another source of uncertainty lies in the choice of a model for absorption from lyman-$\alpha$ system. From observations, Model A2 of Miralda-Escude and Ostriker gives a fair picture of absorption due to lyman-$\alpha$ systems at high redshifts. However, Meiksin \& Madau [32] have argued that absorption due to lyman-$\alpha$ clouds can be smaller (as compared to model A2 of Miralda-Escude and Ostriker). On the other hand, Lanzetta [33] has reported a rapid evolution of lyman-limit systems at high redshifts, which would increase the absorption. However these uncertainties in absorption from lyman-$\alpha$ systems do not change  values calculated using model A2 by more than a factor of few. 

To conclude, we showed that the presence of singly ionized helium in the diffuse IGM can put extremely stringent constraints on radiatively decaying neutrinos. In particular, for $  \tau \ga {\it a \, few}  \times 10^{18} \,{\rm sec}(1 {\rm eV}/m_\nu^2)$, the bounds on radiative lifetime (or equivalently on transtion moment $\mu_{12}$) can be several orders more stringent than the previously known astrophysical and cosmological constraints. More recently, large optical depth from singly ionized helium has been detected in the spectra of two more quasars [34], which would allow one to obtain similar constraints on radiatively decaying neutrinos.  Our results also show that the presence of primordial elements in the IGM  generically give very restrictive  bounds on radiatively decaying neutrinos. 

\bigskip

\beginsection Acknowledgements

The auther would like to thank  A. Goyal, B. Nath, T. Padmanabhan, and P. Pal for useful discussions.

\vfill\eject

\beginsection References

\item{[1]} A. De Rajula , S. L. Glashow , Phys. Rev. Lett. 45, 942 (1980).
\item{[2]} D. W. Sciama ,  Phys. Rev. Lett. 65, 2839 (1990).
\item{[3]} D. W. Sciama, {\it Modern Cosmology and the Dark Matter Problem} (Cambridge University Press, 1993).
\item{[4]} E. W. Kolb and M. S. Turner, {\it The Early Universe} (Addison-Wesley, Redwood City, CA, 1990).
\item{[5]} R. Cowsik, Phys. Rev. Lett. 39, 784 (1977). 
\item{[6]} E. W. Kolb and M. S. Turner, Phys. Rev. Lett. 62, 509 (1989); S. A. Bludman, Phys. Rev. D 45, 4720 (1992);  L. Oberaurer {\etal}, Astropar. Phys. 1, 377 (1993).
\item{[7]} G. G. Raffelt, \apj 365 559 1990 ;V. Castellani and Dengl'Innocenti, \apj 402 574 1993 .
\item{[8]} J. Bernstein and S. Dodelson, Phys. Rev. D 41, 354 (1990);  S. Dodelson and J. M. Jubas, MNRAS 266, 886 (1994).

\item{[9]} J. E. Gunn and B. A. Peterson, \apj 142 1633 1965 .

\item{[10]} E. B. Jenkins and J. P. Ostriker, \apj 376 33 1991 ; E. Giallongo {\etal}, \apj 398 L9 1992 ; J. K. Webb, MNRAS 255, 319 (1991).
\item{[11]} T. M. Tripp, R. F. Green, and J. Bechtold, \apj 364 L29 1990 ; E. A. Beaver {\etal}, \apj 337 L1 1991 ; D. Reimers {\etal}, Nature (London) 360, 561 (1992).
\item{[12]}P. Jakobsen {\etal}, Nature (London), 370, 35 (1994). 
\item{[13]}P. Madau and A. Meiksin, \apj 433 L53 1994 ; A. Songaila, E. L. Hu, and L. L. Cowie, Nature (London) 375, 124  (1995).
\item{[14]} M. Fukugita, Phys. Rev. Lett. 61, 1046 (1988).
\item{[15]} R. Cowsik and J. McClelland, Phys. Rev. Lett. 29, 669 (1972).
\item{[16]} M. S. Turner, Phys. Rev. D 31, 1212 (1985).
\item{[17]} T. Padmanabhan, {\it Structure Formation in the Universe} (Cambridge University Press, 1993).
\item{[18]} J. Miralda-Escud\'e and J. P. Ostriker, \apj 350 1 1990 ; J. Miralda-Escud\'e, MNRAS 262, 273 (1993).
\item{[19]} K. Enqvist, P. Olesen and V. Seminov, Phys. Rev. Lett. 69, 2157 (1992).
\item{[20]} Y. B.  Zeldovich and  R. A. Sunyaev, Asron. Zh. 46, 775 (1969)

\item{[21]} J. C. Mather {\etal}, \apj 420  439 1994 .
\item{[22]} S. Bowyer, ARA \& A, 29, 59 (1991); R. C. Henry, ARA \& A, 29, 89 (1991).
\item{[23]} R. J. Reynolds {\etal}, \apj 309 L9 1986 ; P. Maloney, \apj 414 41 1993 .
\item{[24]} R. L. Brown, \apj 164 387 1971 ; A. A. Zdziarski and R. Svensson, \apj  344 551 1989 .

\item{[25]} S. Bajtik, R. C. Duncan, and J. P. Ostriker, \apj 327 570 1988 .
\item{[26]} W. Marciano and A. I. Sanda, Phys. Lett. B 67, 303, (1977).
\item{[27]} R. N. Mohapatra and P. B. Pal, {\it Massive Neutrinos in Physics and Astrophysics} (World Scientific, Singapore, 1991).
\item{[28]} M. Fukugita and T. Yanagida, in { \it Physics and  Astrophysics of Neutrinos\/} eds.  M.~Fu\-kugita and A. Suzuki (Springer-Verlag, 1994).
\item{[29]} S. K. Sethi, in preparation.
\item{[30]} M. White, G. Gelmini, and J. Silk, Phys. Rev. D 51, 2669(1995)
\item{[31]} B. B. Nath and S. K. Sethi, submitted for publication.
\item{[32]} A. Meiksin and P. Madau, \apj 412 34 1993 .
\item{[33]} K. M. Lanzetta, \apj 375 1 1991 .
\item{[34]} D. Tytler, 1994, unpublished; A. Davidsen, 1995, unpublished.

\vfill\eject

\beginsection Figure Captions

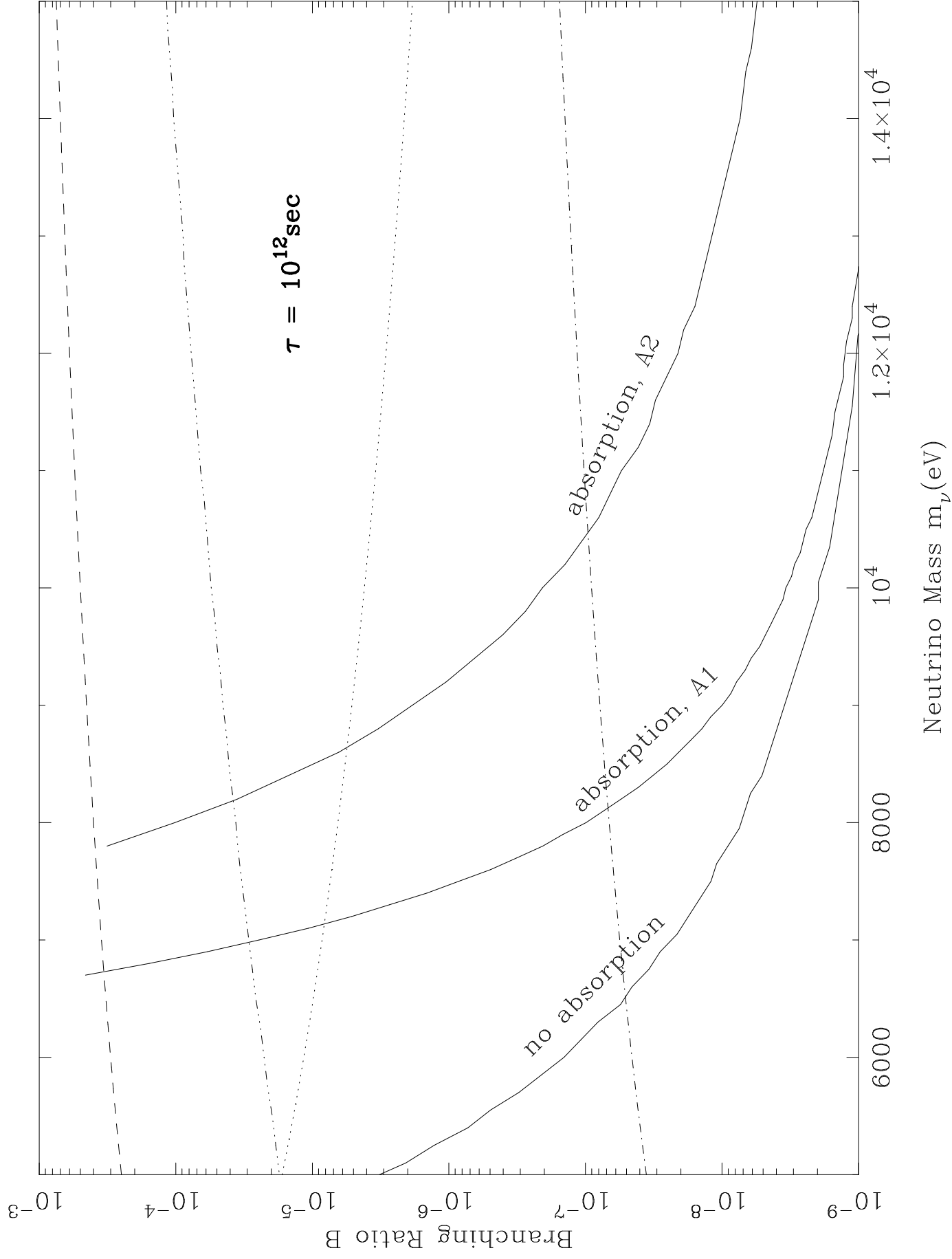
\item{{\bf Figure 1}}:  For $\tau = 10^{12} \, {\rm sec}$,  constraints from singly ionized helium ({\it solid lines} marked with chosen model of absorption  from lyman-$\alpha$ systems) are plotted with constraints from supernova ({\it dashed line}), CBR spectrum ({\it dotted line}),  UV background for absorption model  A2 ({\it dot-dot-dot-dashed line}), and  primordial magentic field ({\it dot-dashed line}). Region above the curves is ruled out by the respective constraints. 

\smallskip

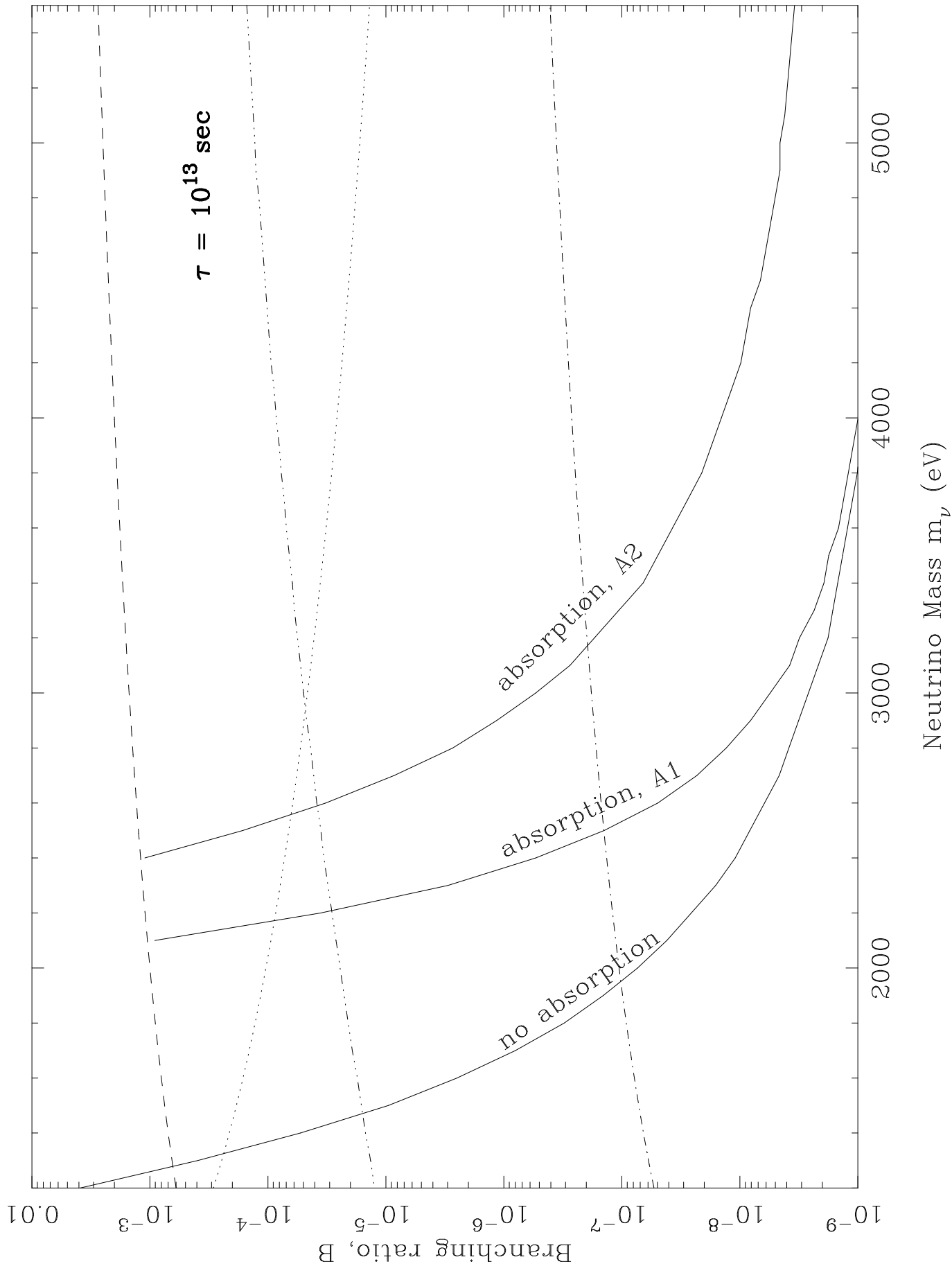
\item{\bf Figure 2}:  Same as fig. (1) but for $\tau = 10^{13}\, {\rm sec} $. 
\smallskip

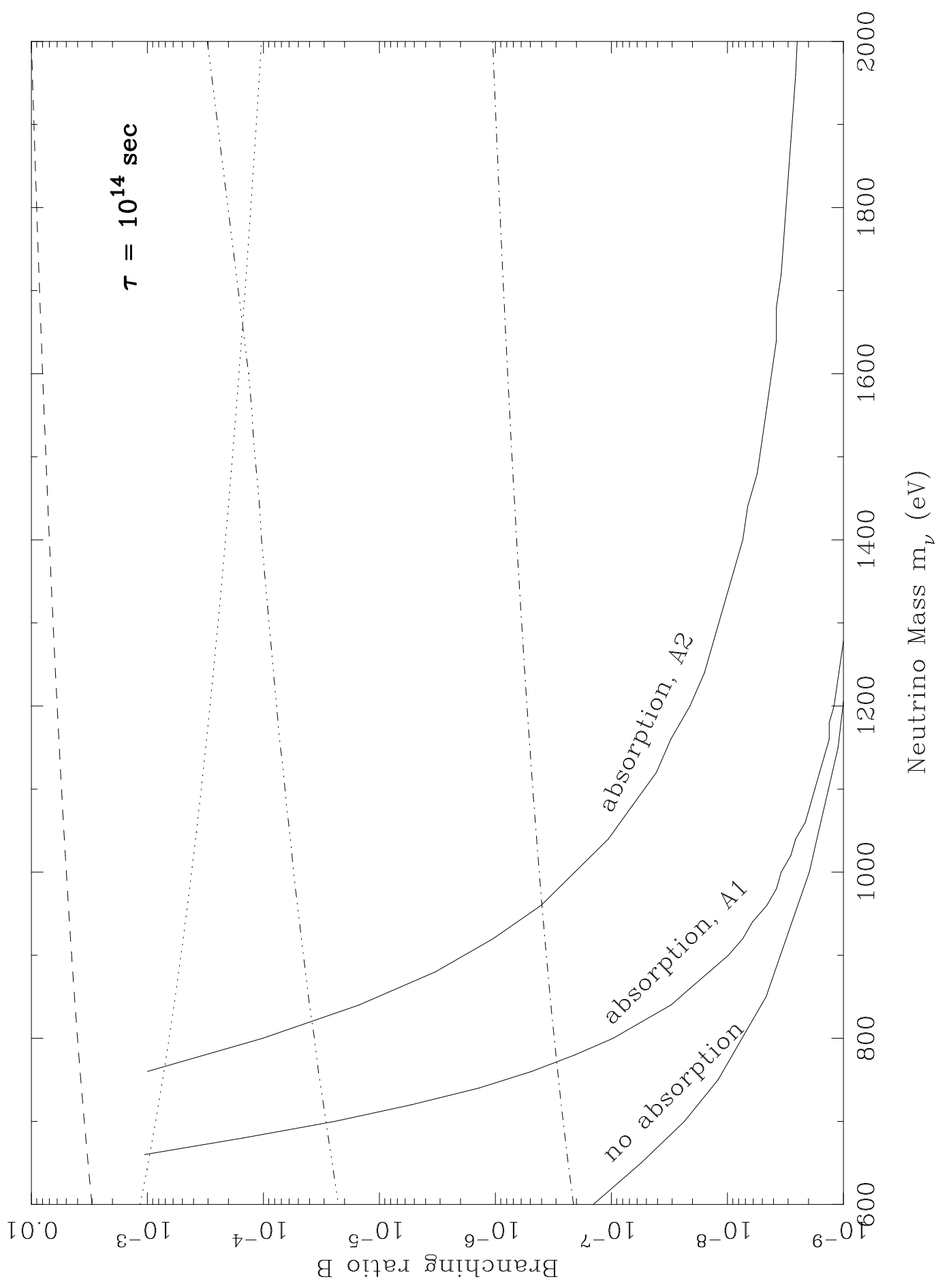
\item{\bf Figure 3}:   Same as fig. (1) but for $\tau = 10^{14} \, {\rm sec} $. 
\smallskip

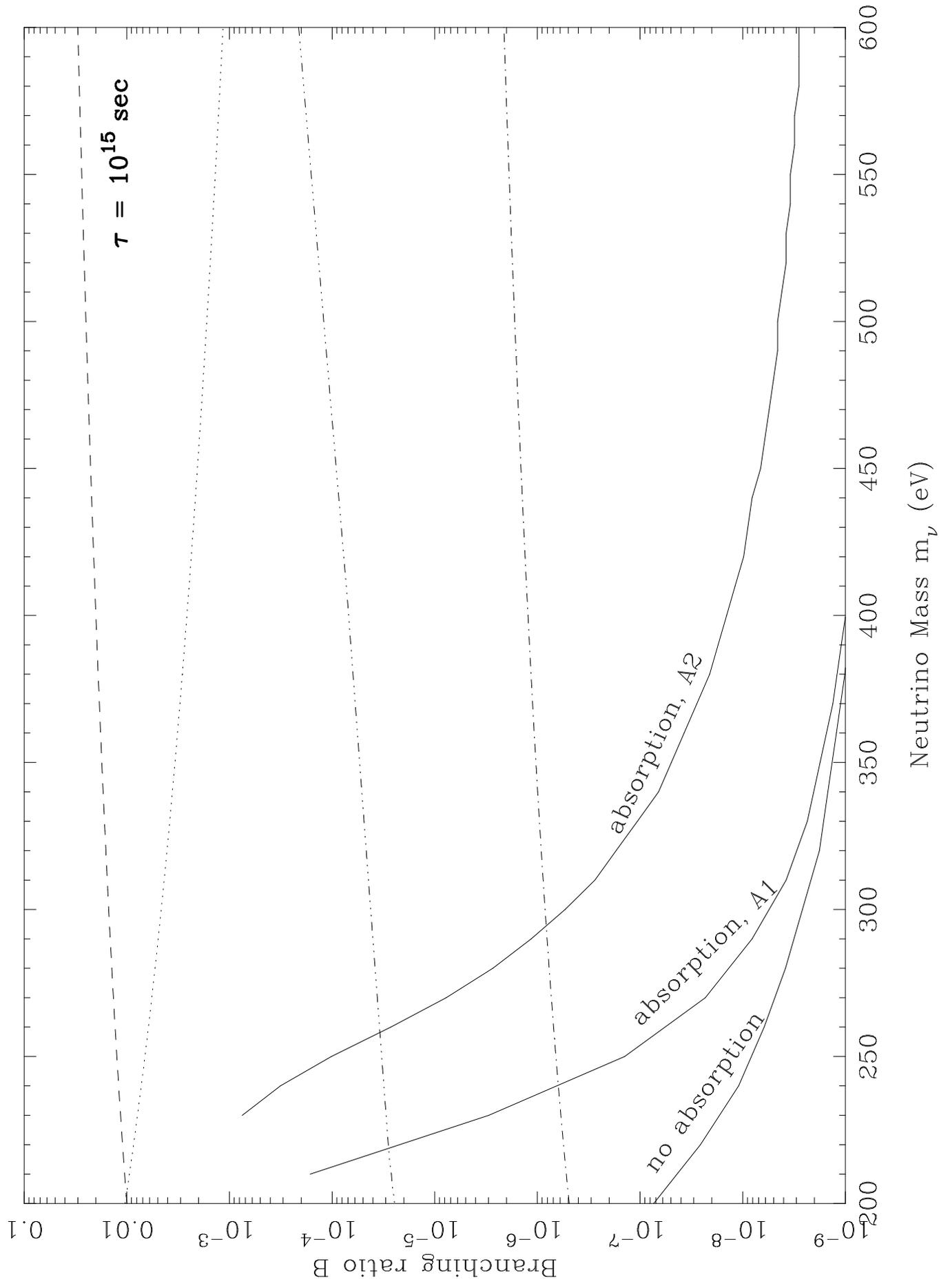
\item{\bf Figure 4}:  Same as fig. (1) but for $\tau = 10^{15} \, {\rm sec}$.

\vfill\eject

\end